\documentstyle[amsmath,amssymb,graphicx]{article}

\def\be{\begin{eqnarray}}
\def\ee{\end{eqnarray}}
\def\nn{\nonumber}

\def\l[{\phantom.[}

\hoffset= - 1.0in         % switch off for draft style
\begin{document}

\hfill ITEP/TH-22/16

\hfill IITP/TH-16/16

\bigskip

\centerline{\Large{
On rectangular HOMFLY for twist knots
 }}

\bigskip

\bigskip

\centerline{{\bf Ya.Kononov$^{d,e,f}$, A.Morozov$^{a,b,c}$}}

\bigskip

{\footnotesize

\centerline{{\small
$^a$ ITEP, Moscow 117218, Russia}}

\centerline{{\small
$^b$ National Research Nuclear University MEPhI, Moscow 115409, Russia
}}

\centerline{{\small
$^c$ Institute for Information Transmission Problems, Moscow 127994, Russia
}}

\centerline{{\small
$^d$ Columbia University, Department of Mathematics, New York, 10027, USA
}}

\centerline{{\small
$^e$ National Research University Higher School of Economics, Moscow, 117312, Russia
}}

\centerline{{\small
$^f$ Landau Institute for Theoretical Physics, Chernogolovka, Russia
}}

}

\bigskip

\bigskip

\centerline{ABSTRACT}

\bigskip

{\footnotesize
As a new step in the study of rectangularly-colored knot polynomials,
we reformulate the prescription of \cite{recttwist}
% arXiv:1606.06015
for twist knots in the double-column representations $R=[rr]$
in terms of skew Schur polynomials.
These, however, are mysteriously shifted from the standard topological locus,
what makes further generalization to arbitrary $R=[r^s]$
not quite straightforward.
}

\bigskip

\bigskip

Knot theory is an old and respected branch of mathematics,
but recently it also became
one of the rapidly developing branches of theoretical physics.
This is because the knot polynomials \cite{knotpols} appeared to
provide  exact {\it non-perturbative} answers
to Wilson-line averages
\be
{\cal H}_R^{\cal K}(A,q) = \left<{\rm Tr}_R\  P\exp\oint_{\cal K} {\cal A} \right>
\label{Wilsonline}
\ee
in $3d$ Chern-Simons theory \cite{CS} --
one of the simplest members of the family of physically relevant Yang-Mills theories.
In (\ref{Wilsonline}) $q$ is made from the coupling constant $k$,
$q = \exp\left(\frac{2\pi i}{k+N}\right)$,
and $A=q^N$ -- from the parameter $N$ of the  gauge group $Sl(N)$.
Remarkably, in these variables the average is a Laurent polynomial -- provided
the space-times is simply-connected.
Despite Chern-Simons is topological theory, i.e. has nearly trivial dynamics
in space-time, dependencies of physical quantities on the other parameters
(coupling constants etc)
are quite non-trivial -- and provide a good model and polygon for
the study of renormalization-group and boundary-condition properties.
Moreover, from this point of view Chern-Simons seems less trivial than, say,
the comprehensible sectors of $N=4$ SYM theory
(in particular, its integrability properties are far more sophisticated) --
still it is exactly solvable, but not yet solved.
Added to this are deep connections of Chern-Simons theories to conformal field theory
and various string models, especially to toric Calabi-Yau compactifications.
The features of knot polynomials are still a set of mysteries,
ranging from a hierarchical set of integrality properties to
various RG-like evolutions in different parameters, especially in the space
of representations $R$, while the standard methods of non-perturbative analysis,
like Ward-identities, AMM/EO topological recursion, integrability techniques
etc are not yet fully applicable.

Development of the theory is still going through consideration of examples:
particular knots ${\cal K}$ and particular representations $R$,
for which a powerful technique is now developed \cite{RT}-\cite{mmms123}.
At present stage these examples start being unified into the simplest families,
either of knots or of representations.
This paper is about a mixture: we provide an exact answer for a one parametric
family of twist knots ${\rm Tw}_m$ in a one-parametric family of two-column rectangular
representations $R=[rr]$.
It is a new small step along the line, originated in \cite{DGR,GGS} and \cite{IMMMfe}-\cite{Konodef}
and recently continued in \cite{rect41,rect41mod} and \cite{recttwist}.
The basic point here is the relative simplicity of {\it differential expansion}
for twisted knots, which allows to {\it guess} answers in big representations
from explicit calculations in the small ones.
Thus the result of this work is a unification of theoretical and experimental
considerations -- what only emphasizes the physical nature of modern knot theory
advances.
Conjectures of \cite{rect41,recttwist} are strongly supported by a recent alternative
calculation in \cite{ShaSle}:
the calculation \cite{mmms123}  of inclusive 3-strand Racah matrices was
extended there to representation $R=[33]$, and so evaluated $[33]$-colored HOMFLY
coincide with the prediction of \cite{recttwist}.

\bigskip

In the present paper we address one of the important claims of \cite{rect41},
which in reformulation of \cite{rect41mod} states that
the rectangular HOMFLY polynomials
for defect-zero knots (those where Alexander polynomial has degree one),
in particular for the twist family ${\rm Tw}_m$,
can be represented as
\be
\boxed{
{\cal H}_{[r^s]}^{(m)}(A,q) =
\sum_{\lambda\subset R} D_{\lambda^{tr}}(r)\cdot D_{\lambda}(s)\cdot Z_{r|s}^\lambda\cdot
{F}^{(m)}_\lambda(A,q)
}
\label{Htw}
\ee
where quantum dimensions
\be
D_{\lambda}(N) := \chi_\lambda\left\{p_k^*\right\}
\ee
are made from the Schur polynomials $\chi_{\lambda/\mu}$ \cite{skewchars} at the topological locus \cite{DMMSS},
\be
p_k^* = \frac{\{A^k\}}{\{q^k\}} =\frac{[Nk]}{[k]}
\ee
we use the standard notation $\{x\} = x-x^{-1}$ and $[k]=\{q^k\}/\{q\}$.
The $Z$-factors, associated with the Young diagrams $\lambda$ are defined as
\be
{ Z}^\lambda_{r|s}(A,q) := \prod_{\square \in \lambda} Z_{r|s}^{(a'(\square) - l'(\square))} =
\prod_{\square \in \lambda} \{A q^{r + a'(\square) - l'(\square)}\} \{A  q^{-s + a'(\square) - l'(\square)}\}
\label{Zlambda}
\ee
and the  dependence on the knot itself is concentrated in the set of {\it polynomial}
factors ${F}^{\cal K}_\lambda(A,q)$, which are instead independent of the
original representation $R=[r^s]$.
The factors $F$ are especially simple for the three simplest twist knots:
figure-eight  ${\cal K}=4_1={\rm Tw}_{-1}$, unknot ${\rm Tw}_0$ and the trefoil $3_1={\rm Tw}_{1}$:
\be
{F}^{4_1}_\lambda={ F}^{(-1)}_\lambda=1 \nn\\
{F}^{unknot}_\lambda = F^{(0)}_{\lambda} = \delta_{\lambda,\emptyset} \nn\\
F^{3_1}_\lambda = F^{(1)}_{\lambda}= (-A^2)^{|\lambda|}\cdot q^{2 \beta(\lambda)}
%\nn \\
%\beta(\lambda) = \sum_{\square \in \lambda} (a'(\Box)-l'(\Box))
\label{sr41}
\ee
with
$\beta(\lambda) = \sum_{\square \in \lambda} (a'(\Box)-l'(\Box))$.

However, for arbitrary twist knots ${\rm Tw}_m$ the factors ${F}^{(m)}_\lambda$
are amusingly non-trivial.
In \cite{evo} they were found for all single-column diagrams $\lambda$ --
and they appeared to be polynomials, composed from the sums of fractions(!).
In \cite{recttwist} the general structure of this decomposition was revealed
and numerous examples were explicitly worked out.
It turned out that the numerators in fractions for multi-column $\lambda$
can also look non-trivial, and it was difficult to work out a general formula
already for the two-column case.
The purpose of the present paper is to resolve {\it this} particular problem:
we recognize in the numerators the skew Schur functions, evaluated at
mysteriously-{\it shifted} topological locus and provide the generic
formula for ${F}^{(m)}_\lambda$ with arbitrary two-column Young diagram $\lambda$:
\be
\boxed{
{F}_\lambda^{(m)} =
\sum_{\mu\subset \lambda} {f}_{\lambda,\mu}\cdot \Lambda_\mu^{2m}
= (-A^2)^{|\lambda|} q^{\beta(\lambda)}
%{{\cal D}_\lambda}
\sum_{\mu\subset \lambda}  (-)^{|\mu|}\cdot
\frac{\chi_{\lambda/\mu}^{**}\chi_{\mu^{tr}}^{**}}{\chi_\lambda^{**}}
\cdot{g}_{\lambda,\mu}\cdot \Lambda_\mu^{2m}
}
\label{Fexpan}
\ee
where sum goes over all the Young sub-diagrams $\mu$ of $\lambda$,
the $m$-dependence is concentrated in the powers of "eigenvalues" $\Lambda_\mu$
and  $g_{\lambda,\mu}(A,q)$ are some ratios of the "differentials"
${D}_k=\{Aq^k\}$.
All non-factorized contributions and even all $q$-number-dependent combinatorial coefficients
are captured by the skew-Schur functions $\chi_{\lambda/\mu}$ \cite{skewchars},
which are defined by decomposition formula
\be
\chi_\lambda\{p'_k+p''_k\} = \sum_{\mu \subset \lambda}
\chi_{\lambda/\mu}\{p_k'\}\cdot \chi_\mu\{p''_k\}
\ee
or
\be
\chi_\lambda\{p'_k-p''_k\} = \sum_{\mu \subset \lambda} (-)^{|\mu|}\,
\chi_{\lambda/\mu}\{p_k'\}\cdot \chi_{\mu^{tr}}\{p''_k\}
\ee
what at $p''=p'=p$ implies
\be
%\sum_{\mu\subset \lambda} K^\lambda_\mu \{p\} =
\boxed{
\sum_{\mu\subset\lambda}  (-)^{|\mu|}\,
\frac{\chi_{\lambda/\mu}\{p\}\,\chi_{\mu^{tr}}\{p\}}{\chi_\lambda\{p\}}
= \frac{\chi_\lambda\{0\}}{\chi_\lambda\{p\}} =  \delta_{\lambda,\emptyset}
}
\label{sr0}
\ee
for non-empty $\lambda$.
Alternatively, skew characters can be expressed through Littlewood-Richardson coefficients:
\be
{\rm if}\ \ \ \ \ \ \ \  \chi_\mu\chi_\nu = \sum_\lambda C^\lambda_{\mu\nu}\chi_\lambda,
\ \ \ \ \ \ \ \ \ \ \ \ \ \ \
{\rm then}\ \ \ \ \ \ \ \   \chi_{\lambda/\mu} = \sum_\nu C^\lambda_{\mu\nu}\chi_\nu
\ee
but this definition is of less use for our purposes.

%In fact, the skew Schur functions are present already in (\ref{Htw}):
%\be
%D_{\lambda^{tr}}(r)\cdot D_{\lambda}(s) = \frac{\chi_{[r^s]/\lambda}^{*}\chi_{\lambda}^*}{\chi_{[r^s]}^*} \frac{\prod_{i \in I} D_i}{\prod_{j \in I'} D_j}
%\label{drs}
%\ee
%However, in (\ref{Fexpan})
The peculiarity of (\ref{Fexpan}) is that characters are taken not at the topological locus
$p_k^* = \{A^k\}/\{q^k\}$, but at the shifted one, $p_k^{**}$ with $A$ multiplied by a $\mu$-dependent
power of $q$, see below.
Eq.(\ref{Fexpan}) is a very general formula, and it presumably holds for arbitrary $\lambda$,
however, explicit expression for the products $f_{\lambda,\mu}$ is currently
available only for the double-column $\lambda$'s (and just a little more).
For expressions like (\ref{Fexpan}) the simple formulas (\ref{sr41}) look like
non-trivial sum rules.
Eq.(\ref{sr0}) provides an archetypical example of such sum rule for skew characters,
but (\ref{sr41}) is its sophisticated deformation, including the $\mu$-dependent shifts
from the topological locus
and ratios of differentials which in this context can be considered as compensating factors.

The sample example is the formula \cite{evo} for pure symmetric representations
where contributing are only the single-column Young diagrams contribute
$\lambda= [a+1]$ and $\mu=[i+1]$,
It can be compactly rewritten through shifted skew characters as
\be
%\boxed{
f_{[a+1],[i+1]} \cdot \frac{\chi_{[a+1]/[i+1]}^{**}\chi_{[i+1]}^{**}}{\chi_{[a+1]}^{**}} \ =\
(-)^{i+1} \ \frac{{D}_{i}!\ {D}_{i-1}!}{{D}_{2i}!}\
\left(\frac{\chi_{[a+1]/[i+1]}^*\chi_{[i+1]}^*}{\chi_{[a+1]}^*}
\right)_{A\longrightarrow A\cdot q^{i+1}} = \nn \\
=  \frac{D_{0}D_{2i+1}}{D_{i+1}D_{i}}
\left(\frac{\chi_{[a+1]/[i+1]}^*\chi_{[i+1]^{tr}}^*}{\chi_{[a+1]}^*}
\right)_{A\longrightarrow A\cdot q^{i+1}}
=  \frac{D_{a}D_{2i+1}}{D_{a+i+1}D_{i}}
\left(\frac{\chi_{[a+1]/[i+1]}^*\chi_{[i+1]^{tr}}^*}{\chi_{[a+1]}^*}
\right)_{A\longrightarrow A\cdot q^i}
\label{f1h}
\ee
Here ${D}_k! = \prod_{i=0}^k D_i = \prod_{i=0}^k \{Aq^i\}$,
with the usual prescription ${D}_{-|k|}! = \prod_{i=1}^{|k|-1} {\cal D}_{-i}^{-1}$,
The ratio of factorials can  be rewritten as
$\frac{{D}_{i}!\ {\cal D}_{i-1}!}{{\cal D}_{2i}!}=
\prod_{k=1}^{i} \frac{\{A^{**}/q^k\}}{\{A^{**}q^k\}}$
with $A^{**} = A\cdot q^{i}$ and can be absorbed in the switch $\chi_{[i+1]} \longrightarrow
\chi_{[i+1]^{tr}} = \chi_{[1^{i+1}]}$.
Dependence of additional factor $\frac{D_{a}D_{2i+1}}{D_{a+i+1}D_{i}}$ at the r.h.s. on $a$
is due to the change of the shift from
$i+1$ to $i$ -- in the former case this particular formula looks simpler, but generalization
to arbitrary 1-hook diagrams $\lambda$ is much better in the latter case:
\be
%\!\!\!\!\!\!\!\!\!\!\!\!\!\!\!\!\!\!\!\!\!\!\!
\!\! \boxed{
F_{(a,b)}^{(m)} = \left(A\,q^{\frac{a-b}{2}}\right)^{a+b+1}
\!\!\!\!\!\!\! \cdot \frac{D_0}{D_a!\,\bar D_b!} \cdot
 \left(1 + \sum_{i=0}^a\sum_{j=0}^b G_{(a,b)}^{(i,j)}
%  \ \ \ \ \ \ \ \ \ \ \ \ \ \ \ \ \ \ \ \ \ \ \ \ \ \ \ \ \ \ \ \ \  \nn \\
% \left. + \sum_{i=0}^a\sum_{j=0}^b
%\frac{D_a!\,D_{a+i-j}!}{D_{a+i+1}!\,D_{a-j-1}!}\cdot \frac{\bar D_b!\,\bar D_{b-i+j}!}{\bar D_{b+j+1}!\,\bar D_{b-i-1}!}
%\cdot \frac{D_{2i+1}\,\bar D_{2j+1}}{D_{i-j}^2}
\cdot(A\, q^{i-j})^{2m(i+j+1)}\cdot
\left(\frac{\chi_{(a,b)/(i,j)}^*\chi_{(i,j)^{tr}}^*}{\chi_{(a,b)}^*}
\right)_{\!\!A\longrightarrow A\cdot q^{i-j}}
\right)
}
\label{F1h}
\ee
with
\be
G_{(a,b)}^{(i,j)} = \frac{D_a!\,D_{a+i-j}!}{D_{a+i+1}!\,D_{a-j-1}!}\cdot \frac{\bar D_b!\,\bar D_{b-i+j}!}{\bar D_{b+j+1}!\,\bar D_{b-i-1}!}
\cdot \frac{D_{2i+1}\,\bar D_{2j+1}}{D_{i-j}^2}
\ee
It is important that (\ref{F1h}) is symmetric under the change $(a,i,q)\longleftrightarrow (b,j,-1/q)$,
and this requires the shift to be $i-j$, what means $i$ rather than $i+1$ at $j=0$.

\bigskip

The first unity in brackets in (\ref{F1h}) describes the contribution of the empty sub-diagram
$\mu=\emptyset$, and it is different from all other contributions.
The situation will be similar for multi-hook $\lambda$:
there will be different series of terms, associated with different number of hooks --
and even arms and legs -- in the sub-diagrams $\mu$.
For the two-column $\lambda=[r_1,r_2]$,
there are three different classes of non-empty $\mu\subset\lambda$,
which can be pictorially represented as

\begin{picture}(500,170)(-50,-130)
\put(0,0){\line(1,0){30}}
\put(0,0){\line(0,-1){100}}
\put(15,-15){\line(1,0){15}}
\put(15,-15){\line(0,-1){85}}
\put(30,0){\line(0,-1){70}}
\put(0,-100){\line(1,0){15}}
\put(15,-70){\line(1,0){15}}
\put(7.5,-7.5){\line(0,-1){70}}
\put(-40,-115){\mbox{$\mu=[i_1+1]=(i_1,0||\emptyset)$}}
\put(150,0){
\put(-70,20){\mbox{$\lambda = [r_1,r_2]=[a_1+1,a_2+2]=(a_1,1||a_2,0)$}}
\put(0,0){\line(1,0){30}}
\put(0,0){\line(0,-1){100}}
\put(15,-15){\line(1,0){15}}
\put(15,-15){\line(0,-1){85}}
\put(30,0){\line(0,-1){70}}
\put(0,-100){\line(1,0){15}}
\put(15,-70){\line(1,0){15}}
\put(7.5,-7.5){\line(0,-1){70}}
\put(7.5,-7.5){\line(1,0){17}}
\put(-50,-115){\mbox{$\mu=[i_1+1,1]=(i_1,1||\emptyset)$}}
}
\put(300,0){
\put(0,0){\line(1,0){30}}
\put(0,0){\line(0,-1){100}}
\put(15,-15){\line(1,0){15}}
\put(15,-15){\line(0,-1){85}}
\put(30,0){\line(0,-1){70}}
\put(0,-100){\line(1,0){15}}
\put(15,-70){\line(1,0){15}}
\put(7.5,-7.5){\line(0,-1){70}}
\put(7.5,-7.5){\line(1,0){17}}
\put(22.5,-22.5){\line(0,-1){22.5}}
\put(-60,-115){\mbox{$\mu=[i_1+1,i_2+2]=(i_1,1||i_2,0)$}}
}
\end{picture}

\noindent
where $a$ and $i$ refer to the "pyramid" notation
$\lambda = (a_1,b_1||a_2,b_2||\ldots)$ and $\mu = (i_1,j_1||i_2,j_2||\ldots)$
of \cite{rect41} and \cite{recttwist}
(this is actually the Frobenius parametrization of Young diagrams by hook variables).
In this notation $a_{f+1}\leq a_f$, $i_{f+1}<i_f$, $i_f\leq a_f$ and similarly for $b$ and $j$.
Note that $a_f=b_f=0$ correspond to single-box floor/hook, not to an empty one.

The building blocks for the functions $f_{\lambda,\mu}^{(m)}= f_{\lambda}^{\mu}\cdot\Lambda_\mu^{2m}$
will be denoted by $g_\lambda^\mu$:
\be
g^\emptyset_{(a,b)} := \frac{1}{\prod_{k=-b}^a D_k} = \frac{D_0}{{ D}_a!\,\bar{D}_b!} \nn \\
g^{(i,j)}_{(a,b)} := (-)^{i+j+1}\cdot (A\cdot q^{i-j})^{2m(i+j+1)}\cdot
\frac{[a]!}{[a-i]![i]!}\cdot\frac{[b]!}{[b-j]![j]!}\cdot \frac{[a+b+1]}{[i+j+1]}\cdot
 \frac{{ D}_i!\, \bar{D}_{j}!}{{ D}_{a+i+1}!\,\bar{ D}_{b+j+1}!}\cdot
\frac{D_{2i+1} \bar D_{2j+1}}  {D_{i-j}}
\ee
We absorbed $m$-dependence into $g$, but suppressed this in the notation to make formulas readable.

In terms of $g$-functions for the 1-hook $\lambda=(a,b)$
\be
F_{(a,b)}^{(m)} = \left(Aq^{\frac{a-b}{2}}\right)^{a+b+1}\cdot \left(g_{(a,b)}^\emptyset
+ \ \sum_{i=0}^a \sum_{j=0}^b g_{(a,b)}^{(i,j)}
%\cdot (A\cdot q^{i-j})^{2m(i+j+1)}
\right)
\ee
This is the same quantity as (\ref{F1h}) and comparison explains how $g$ are expressed through the
skew characters.
Likewise in the 2-hook case we have for $\lambda=(a_1,b_1||a_2,b_2)$
\be
F_{(a_1,b_1||a_2,b_2)}^{(m)} = \prod_{f=1}^2 \left(Aq^{\frac{a_f-b_f}{2}}\right)^{a_f+b_f+1}
\cdot \left(g_{(a_2,b_2)}^\emptyset\cdot g_{(a_1,b_1)}^\emptyset
+ \ g_{(a_2,b_2)}^\emptyset\cdot \sum_{i_1=0}^{a_1} \sum_{j_1=0}^{b_1} g_{(a_1,b_1)}^{(i_1,j_1)}\cdot
\xi_{(a_1,b_1||a_2,b_2)}^{(i_1,j_1||\emptyset)}
+ \right.\nn \\ \left.
+ \sum_{i_1=0}^{a_1} \sum_{i_2=0}^{{\rm min}(a_2,i_1-1)}
\sum_{j_1=0}^{b_1} \sum_{j_2=0}^{{\rm min}(b_2,j_1-1)}
g_{(a_2,b_2)}^{(i_2,j_2)}\cdot g_{(a_1,b_1)}^{(i_1,j_1)}
\cdot  {\frac{D_{i_1+i_2+1}\bar D_{j_1+j_2+1}}{D_{i_1-j_2}\bar D_{j_1-i_2}}}\cdot
\xi_{(a_1,b_1||a_2,b_2)}^{(i_1,j_1||i_2,j_2)}
\right)
\ee
$m$-dependence is hidden in $g$-functions,
the three terms in the sum correspond to three cases in the picture (where $b_1=1$ and $b_2=0$),
and $m$-independent correction factors $\xi_\lambda^\mu$ are expressed through the skew-Schur functions.
The main result of this paper is explicit formula for these factors,
restricted to the case $b_2=0$ (constraint $b_1=1$, imposed in the pictures is actually relaxed):

\be
\boxed{\begin{array}{c}
\\
\xi_{(a_1,b_1||a_2,b_2)}^{(0,0 )} = \frac{D_{a_2}\bar D_{b_2}}{D_{a_2+1}\bar D_{b_2+1}}\cdot
%\hat S_{q^{i_1}}
\left(\frac{K_{(a_1,b_1||a_2,b_2)}^{(0,0 )}}{K_{(a_1,b_1 )}^{(0,0)}}
\right)_{A\longrightarrow A }  \\  \\
\xi_{(a_1,b_1||a_2,b_2)}^{(i_1,0 )} = \frac{D_{a_2-0}\bar D_{b_2-i_1}}{D_{a_2+i_1+1}\bar D_{b_2+0+1}}\cdot
%\hat S_{q^{i_1}}
\left(\frac{K_{(a_1,b_1||a_2,b_2)}^{(i_1,0 )}}{K_{(a_1,b_1 )}^{(i_1,0)}}
\right)_{A\longrightarrow A\cdot q^{i_1}}  \\  \\
\xi_{(a_1,b_1||a_2,b_2)}^{(0,j_1 )} = \frac{D_{a_2-j_1}\bar D_{b_2}}{D_{a_2+0+1}\bar D_{b_2+j_1+1}}\cdot
%\hat S_{q^{i_1}}
\left(\frac{K_{(a_1,b_1||a_2,b_2)}^{(0,j_1 )}}{K_{(a_1,b_1 )}^{(0,j_1)}}
\right)_{A\longrightarrow A\cdot q^{-j_1}}  \\  \\
\hline
\\
i_1,j_1\geq 1: \ \ \ \ \ \ \xi_{(a_1,b_1||a_2,0)}^{(i_1,j_1 )} =
%\hat S_{q^{i_1+1}}
\left(\frac{K_{(a_1,b_1||a_2,0)}^{(i_1,j_1) }}{K_{(a_1,b_1) }^{(i_1,j_1)}}
\right)_{A\longrightarrow A\cdot q^{i_1+1}}  \\ \\
\xi_{(a_1,b_1||a_2,0)}^{(i_1,j_1||i_2,0) }
= \left(
\frac{K_{(a_1,b_1||a_2,0)}^{(i_1,j_1||i_2,0)}}{K_{(a_1,b_1 )}^{(i_1,j_1 )}
\cdot K_{(a_2,0 )}^{(i_2,0 )}}
\right)_{A\longrightarrow A\cdot q^{i_1+i_2+2}} \\
\end{array} }
\label{b1b2}
\ee

\bigskip

\noindent
Here $K_\lambda^\mu $ can be either $ \frac{\chi_{\lambda/\mu}^*\cdot \chi_\mu^*}{\chi_\lambda^*}$
or $ \frac{\chi_{\lambda/\mu}^*\cdot \chi_{\mu^{tr}}^*}{\chi_\lambda^*}$ --
the difference between $\chi_\mu^*$ and $\chi_{\mu^{tr}}^*$ drops away from the ratios.
The first three of these formulas have good chances to be true for all $b_2\geq 0$.
However, in the last two formulas the shifts  do not respect the symmetry,
associating transposition of the diagrams $a_f\leftrightarrow b_f$, $i_f\leftrightarrow j_f$
with the change $q\leftrightarrow -q^{-1}$ -- thus they can {\it not} be true for arbitrary $b_2$.
Instead for the double-column $\lambda$, i.e. for $b_1=1,b_2=0$,  the shifts in all these formulas
are by $|\mu|-1$ -- the only exception
is in the third line, but $b_1=1$ allows only $\mu=(0,1)$ there, and this is the case of full
factorization, when shifts do not matter.

Anyhow, at $b_2=0$ and arbitrary $b_1$ eqs.(\ref{b1b2}) work perfectly well as they are:
one can check that these formulas provide {\it polynomials} for $F^{(m)}_\lambda$
at arbitrary $\lambda$ and $m$ -- and satisfy the necessary sum rules (\ref{sr41}) at $m=-1,0,1$.

\bigskip

These formulas are sufficient to describe HOMFLY for twist knots in arbitrary
double-column representations $R=[rr]$
and -- by the change $q\longrightarrow -q^{-1}$ -- in arbitrary double-line $R=[2^r]$,
thus providing the generalization of the result of \cite{evo} for arbitrary symmetric $R=[r]$ and
antisymmetric $R=[1^r]$.
Generalization to superpolynomials \cite{sup} and application to Racah calculus {\it a la}
\cite{inds,pret,mmms123}
are straightforward --
along the lines of \cite{rect41mod} and \cite{recttwist} respectively.
Reformulation in terms of shifted skew characters seems to resolve the main {\it technical} puzzle of
\cite{recttwist}.
The answer of \cite{recttwist} for exclusive unitary Racah matrix $\bar S$
is now understood to be bilinear in shifted skew characters:
\be
 \bar S_{\mu\nu} = \frac{D_R}{{\cal D}_\mu {\cal D}_\nu}
\sum_{\mu,\nu\subset \lambda \subset R=[r^s]}
\frac{D_{\tilde\lambda}(r)\cdot D_\lambda(s)\cdot Z_{r|s}^{\lambda}}{F_\lambda^{(-1)}(q,A)}\cdot
f_{\lambda,\mu}\cdot f_{\lambda,\nu}
\sim \sum_{\lambda\in [r^s]} C_\lambda\cdot \chi_{\lambda/\mu}^{**}\cdot \chi_{\lambda/\nu}^{**}
\ee
with ${\cal D}_\mu$ dimension of representation in $R\otimes \bar R$,
associated with $\mu \subset R$.
The weights $C_\lambda$ are fully known from (\ref{b1b2}) for the case of $R=[rr]$.
The second exclusive Racah matrix $S$   diagonalizes this $\bar S$ by the usual rule \cite{bSSS}
\be
\bar T \bar S \bar T = S T^{-1} S^\dagger
\ee
where $\bar T = {\rm diag}(\Lambda_\mu)$ and $T$ is another diagonal matrix, made from
the eigenvalues of $\bar S$ (they are actually the ${\cal R}$-matrix eigenvalues
in the channel $R\otimes R$).
Thus $S$ can be calculated from a known $\bar S$ for every particular representation $R=[rr]$.
These  formulas, however, can not be  the end of the story,
and should possess further simplifications, also making transparent the unitarity of $\bar S$ and $S$.
In their present form they only add to conceptual mystery about the origin of  $F$-factors and
a variety of associated sum rules.
Also unclarified remains the growth of complexity in formulas with the increasing number of hooks
and its relation to a somewhat similar phenomenon for Alexander polynomials \cite{DMMSS,Alex}.
We are going to address these issues in further publications.

\section*{Acknowledgements}

Our work is partly supported by RFBR grants 16-01-00291 (Y.K.), 16-02-01021 (A.M.)
by young scientist grants 16-31-00484 (Y.K.), 15-31-20832-mol-a-ved (A.M.),
by the Russian Academic Excellence Project
'5-100' and Simons Foundation (Y.K.), by the joint grants
15-51-52031-HHC, 15-52-50041-YaF, 16-51-53034-GFEN, 16-51-45029-Ind.

\end{document}